\input amstex
\input vanilla.sty
\pagewidth{15.5 cm}
\pageheight{22.8 cm}
\magnification=\magstephalf
\TagsOnRight
\baselineskip 16.2pt
\define \la {\lambda}
\define \part {\partial}
\centerline {\bf Binary
Nonlinearization of Lax pairs of Kaup-Newell Soliton Hierarchy}

\vskip 4mm
\centerline{Wen-Xiu Ma$^{\dagger\ddagger}$, Qing Ding$^\dagger$, 
Wei-Guo Zhang$^*$
and Bao-Qun Lu$^\dagger$}

\vskip 3mm
\line 
{\hbox to 0pt {\hss}
\hss $\displaystyle
\matrix \format \l \\
\text{{\it $^\dagger$Institute of Mathematics, Fudan University, 
Shanghai 200433, 
P. R. of China}}\\
\text{{\it $^\ddagger$FB Mathematik und Informatik, Universit\"at Paderborn,
D-33098 Paderborn, Germany}}
\\
\text{{\it $^*$Dept. of Math. $\&$ Mech., Changsha Railway University, 
Changsha 410075, 
P. R. of China}}
\endmatrix
$\hss
\hbox to 0pt {\hss}}

\vskip5mm

\vskip3mm

{\narrower 
{\bf Summary.} --- Kaup-Newell soliton hierarchy is derived from
a kind of Lax pairs different from the original ones.
Binary nonlinearization procedure corresponding to the Bargmann
symmetry constraint 
is carried out for those Lax pairs.
The proposed Lax pairs together with
adjoint Lax pairs are constrained 
as a hierarchy of
commutative, finite dimensional integrable Hamiltonian systems in the 
Liouville sense, which also provides us with new examples of 
finite dimensional integrable Hamiltonian systems.
A sort of  involutive 
solutions to the Kaup-Newell hierarchy are exhibited through the obtained
finite dimensional integrable systems and
the general involutive system engendered by binary nonlinearization
is reduced to a specific involutive
system generated by mono-nonlinearization.

PACS 03.40K - waves and wave propagation: general mathematical aspects
\par} 

\vskip6mm
\centerline{{\bf 1. Introduction}}

\vskip2mm

It is very important (and interesting) 
to search for integrable reductions of
soliton equations. Similarity reductions lead to
equations of Painlev\'e type [1], and algebraic pole expansion
solutions to soliton equations engender finite dimensional integrable 
$N$-body Hamiltonian systems [2]. These are two main examples of integrable 
reductions.

Recently a systematic approach called the nonlinearization (more
precisely, we call it mono-nonlinearization) of Lax pairs is introduced 
for generating finite dimensional completely
integrable systems from soliton hierarchies by Cao and Geng et al.
[3-5]. Based upon the
idea of nonlinearization, one of the authors proposed
binary nonlinearization technique for soliton hierarchies [6,7],
which was successfully applied to a few well-known soliton
hierarchies [6,7].
Nonlinearization method may yield a
large variety of finite dimensional 
completely integrable Hamiltonian systems whose
independent integrals of motion can be explicitly given out, and it
provides a way of solving soliton equations by separation of spatial and 
temporal variables. 

The integrable reductions in the mono-nonlinearization and
binary nonlinearization are related to a kind of specific symmetry
constraints [6,7]. Bargmann symmetry constraints
(usually the first order with respect to derivative)
lead to the most interesting reductions in the nonlinearization
because of their simplicity.
 These constraints require conserved covariants to be linear functions
but not differential functions with respect to the potentials.
The higher order symmetry constraints need to introduce
new dependent variables,
the so-called Jacobi-Ostrogradsky coordinates.
The spatial parts of the resulting nonlinearized Lax pairs
by nonlinearization 
are exactly the restricted flows named by Antonowicz and Wojciechowski
etc. [8].  

In this paper, we would like to discuss
binary nonlinearization  for Kaup-Newell 
hierarchy [9] of soliton equations.
In Section 2, following the standard procedure, we shall reconstruct
Kaup-Newell soliton hierarchy from a particular spectral problem
other than the original Kaup-Newell spectral problem [9] and
give rise to some algebraic properties connected with Kaup-Newell hierarchy. 
In Section 3, we shall in detail 
exhibit binary nonlinearization procedure corresponding to the Bargmann
constraint with zero boundary conditions and without 
zero boundary conditions and vigorously constrain the proposed  Lax pairs and
adjoint Lax pairs of Kaup-Newell hierarchy into a hierarchy of
commutative, finite dimensional integrable Hamiltonian systems 
in the Liouville sense.
By the resulting finite dimensional
integrable system, we shall further exhibit a sort of involutive 
solutions to Kaup-Newell hierarchy.
Finally in Section 4, we reduce
the general involutive system derived from binary nonlinearization
to a specific involutive
system generated by mono-nonlinearization and give some other remarks. 

\vskip5mm

\centerline{{\bf 2. New Lax pairs of Kaup-Newell hierarchy and 
some related properties}}

\vskip2mm

Let us consider the following specific spectral problem
$$\phi _x=U\phi =U(u,\la )\phi=\pmatrix \la &q\\ \la r & -\la \endpmatrix 
\phi,\ \phi = \pmatrix \phi _1 \\ \phi _2 \endpmatrix ,\ u =
\pmatrix q \\ r \endpmatrix.\tag2.1$$
This spectral problem is a little different from the original Kaup-Newell 
spectral problem [9].
We find by a large amount of computation that it is a possible 
spectral problem leading 
to an integrable hierarchy. Although we don't know if there exists a gauge 
transform between (2.1) and the original spectral problem,
we  shall see later that its related integrable
systems may be transformed into
Kaup-Newell hierarchy.

In order to derive isospectral systems associated with (2.1),
similar to Ref. [10]
we first solve the  
adjoint representation equation (see Ref. [11]) $V_x=[U,V]$ of
$\phi _x=U\phi$.
Take $$V =\pmatrix a&\la ^{-1}b\\c&-a\endpmatrix 
.$$ Noting that 
$$[U,V]=\pmatrix qc-rb&2b-2qa\\2\la ra -2\la c&rb-qc\endpmatrix,$$
we find that the adjoint representation equation $V_x=[U,V]$ becomes 
$$\left \{\aligned &a_x= q c-r b ,\\ & \la ^{-1}b_x=2 b-2 q a , \\ & c_x=
2\la ra -2\la c.\endaligned \right.\tag2.2$$  On setting 
$a=\sum_{i\ge0}a_i\la ^{-i},\, b=\sum_{i\ge0}b_i\la ^{-i},\,
c=\sum_{i\ge0}c_i\la ^{-i},$ (2.2) yields equivalently 
$$\left \{ \aligned  &2b_0-2qa_0=2ra_0-2c_0=0,\\& a_{ix}=qc_{i}-r b_{i}, \\ 
&b_{ix}=2b_{i+1}-2qa_{i+1}, \qquad i\ge 0. \\  &c_{ix}=2ra_{i+1}-2c_{i+1},
\endaligned \right.   \tag2.3$$
We choose  $a_0=1,\ b_0=q,\ c_0=r$ and assume that $a_i|_{[u]=0}=
b_i|_{[u]=0}= c_i|_{[u]=0}=0,\ i\ge 1$, where $[u]=(u,u_x,\cdots)$
 (this means to select zero 
constants for integration).
Because we have  $$\align &2a_{i+1,x}=2qc_{i+1}-2rb_{i+1}\\
&=q(2ra_{i+1}-c_{ix})-r(b_{ix}+2qa_{i+1})\\ &= -qc_{ix}-rb_{ix} ,\endalign$$
the equality (2.3) gives rise to  a recursion relation for determining 
$a_i,\,b_i,\,c_i$:
$$\left \{\aligned  &a_{i+1}=-\frac12 \part ^{-1}(qc_{ix}+rb_{ix}),\\
&b_{i+1}=\frac12b_{ix}-\frac12 q\part ^{-1}( qc_{ix}+rb_{ix}),\ i\ge0.\\
&c_{i+1}=-\frac12c_{ix}-\frac12 r\part ^{-1}( qc_{ix}+rb_{ix}),
\endaligned \right.\tag2.4$$ 
Therefore, for instance, we can obtain 
$$\align & a_1=-\frac12 qr,\  b_1=\frac12(q_x-q^2r),
\ c_1=-\frac12 (r_x+qr^2);\\
& a_2=\frac38 q^2r^2+\frac14 qr_x-\frac14q_xr,\ 
b_2=\frac14 q_{xx}-\frac34 qq_xr+\frac38q^3r^2,\ 
c_2=\frac14 r_{xx}+\frac34 qrr_x+\frac38q^2r^3.  \endalign$$
From $(V^2)_x=[U, V^2]$, we see that
$(\frac 12 \text{tr}V^2)_x=(a^2+\la ^{-1}bc)_x=0$. Thus
we have $a^2+\la ^{-1}bc=1$ by observing $ (\frac 12 \text{tr}V^2)|_{[u]=0}=1$.
Further we may obtain
$$2a_m=-\sum _{i=1}^{m-1}a_ia_{m-i}-\sum_{i=0}^{m-1}b_ic_{m-1-i},\ m\ge2.
\tag2.5$$
This equality shows by an argument of mathematical induction 
that $a_i,b_i,c_i$ are all differential polynomial functions of $u$.

Now we introduce the following 
auxiliary problem associated with the spectral problem (2.1) 
$$\phi _{t_m}=V^{(m)}\phi ,\ V^{(m)}=(\la ^{m+1}V)_+-\pmatrix a_{m+1}&0\\
c_{m+1}&-a_{m+1}\endpmatrix=\pmatrix \la (\la ^m a)_+&(\la ^m b)_+\\
\la (\la ^m c)_+&-\la (\la ^m a)_+\endpmatrix,\ m\ge 0,\tag2.6$$
where the symbol $+$ stands for the selection of the polynomial part of
$\la $. Then cross differentiation (i.e. taking $\part /\part t_m$ (2.1), $
\part /\part x$ (2.6) and setting them equal)
leads to  isospectral integrable systems
$$u_{t_m}= \pmatrix q\\r\endpmatrix_{t_m}=K_m=
\pmatrix b_{mx}\\c_{mx}\endpmatrix  =\Phi ^m u_x,\ m\ge 0.\tag2.7$$
Here the operator $\Phi$ reads
$$\Phi= \pmatrix \frac12 \part-\frac12 \part q\part ^{-1}r&-\frac12
\part q\part ^{-1}q\\ \vspace {2mm} -\frac12 \part r\part ^{-1}r& 
-\frac12 \part
-\frac12 \part r\part ^{-1}q\endpmatrix ,\ \part =\frac \part {\part x},$$
which may be verified to be a hereditary symmetry operator [12],
i.e. to satisfy 
$$\Phi ^2[K,S]+[\Phi K,\Phi S]-\Phi \{ [K,\Phi S]+[\Phi K,S]\}=0,$$
for arbitrary vector fields $K,S$. 
Obviously, under $\frac \part{\part x}\rightarrow  -i
\frac \part{\part x},\ 
\frac \part{\part t_m}\rightarrow  -i
\frac \part{\part t_m},$ (2.7) is transformed into Kaup-Newell
hierarchy [9]. We call (2.7) as a kind of real form of 
Kaup-Newell hierarchy because (2.7) itself is a hierarchy of real systems.
The first two
nonlinear systems in the hierarchy (2.7) are as follows
$$\left \{\aligned &q_{t_1}=\frac 12 \bigl[q_{xx}-(q^2r)_x\bigr],\\
& r_{t_1}=-\frac 12\bigl[ r_{xx}+(qr^2)_x\bigr];\endaligned \right.
\left \{\aligned &q_{t_2}=\frac 14q_{xxx}-\frac34
(qq_xr)_x+\frac38(q^3r^2)_x,\\ &r_{t_2}=\frac 14r_{xxx}+\frac34
(qrr_x)_x+\frac38(q^2r^3)_x. \endaligned \right.$$
The former system may be rewritten as the derivative nonlinear
Schr\"odinger equation [9]  
$$iq_{t_1}=-\frac12 q_{xx}+\frac12 i(q^*q^2)_x$$
under 
$\frac \part{\part x}\rightarrow  -i
\frac \part{\part x},\ 
\frac \part{\part t_1}\rightarrow  -i
\frac \part{\part t_1},$ $q=-r^*$. 

We below want to tersely exhibit bi-Hamiltonian structures of Kaup-Newell
hierarchy (2.7) by applying a powerful tool, i.e. the so-called trace identity
proposed in Ref. [13,14]. Towards this end,
we need, as is usual, the following quantities which are easy to work out: 
$$ <V, \frac {\part U} {\part \la}>= 2a+\la ^{-1}rb,\ 
<V, \frac {\part U} {\part q}>= c,\  <V, \frac {\part U} {\part r}>= b, $$
where $<\cdot,\cdot>$ denotes the Killing form: $<A,B>=\text{tr}(AB)$. 
Then by means of the trace identity [13]
$$\frac \delta {\delta u} <V, \frac {\part U} {\part \la}>= \la^{-\gamma}
\frac \part {\part \la}\la^\gamma ( <V, \frac {\part U} {\part q}>,
<V, \frac {\part U} {\part r}> )^T,\ \gamma =\text{const.},$$
we obtain immediately
$$\frac \delta {\delta u}(2a+\la ^{-1}rb)=
\la^{-\gamma } \frac \part {\part \la}\la^\gamma (c,b)^T.$$
Comparing the coefficients of $\la^{-m-1}$ on two sides of the
above equality and noticing the equality (2.3), we have
$$\frac \delta {\delta u}\bigl[2a_{m+1}+\frac12 (rb_{m}+qc_m)\bigr]=
\frac \delta {\delta u}(2a_{m+1}+rb_{m})=
(\gamma -m)(c_m,b_m)^T,\  m\ge0.\tag2.8 $$
The equality (2.8) with $m=0$ yields  that the constant $\gamma =0$. 
Therefore (2.8) gives rise to an important  formula
$$\frac \delta {\delta u}\bigl[2a_{m+1}+\frac12 (rb_{m}+qc_m)\bigr]=
 -m(c_m,b_m)^T,\  m\ge0.\tag2.9 $$
At this stage, Kaup-Newell hierarchy (2.7) may be transformed into the
Hamiltonian form $$u_{t_m}= \pmatrix q\\r\endpmatrix_{t_m}=K_m=
\pmatrix b_{mx}\\c_{mx}\endpmatrix= JG_m=J\Psi^m
\pmatrix c_0\\b_0\endpmatrix=J 
\frac {\delta H_m}{\delta u} ,\ m\ge 0.\tag2.10$$
Here the Hamiltonian operator $J$, the recursive operator $\Psi$ and
the Hamiltonian functions $H_m$ are defined by $$ \align &
J= \pmatrix 0&\part \\ \vspace {2mm}\part &0\endpmatrix,\ \Psi=\Phi ^*
=\pmatrix  -\frac12 \part -\frac12 r\part ^{-1} q\part &
-\frac12 r\part ^{-1}r\part  \\ \vspace {2mm}
-\frac12 q\part ^{-1}q\part  &\frac12 \part 
-\frac12 q\part ^{-1}r\part  \endpmatrix,\tag2.11a\\ &
H_0=-2a_1,\ H_m=-\frac1 m \bigl[2a_{m+1}+\frac12 (rb_m+qc_m)\bigr]
,\  m\ge1. \tag2.11b \endalign$$
A direct calculation can show that $J$ and $J\Psi$
constitutes a Hamiltonian pair. Therefore the flows determined 
by Kaup-Newell hierarchy commute with each other, and  every Kaup-Newell 
system has an infinite number of symmetries
$\{K_n\}_{n=0}^\infty$ and conserved quantities $\{H_n\}_{n=0}^\infty$.
It is known that  Magri [15] 
showed the existence of bi-Hamiltonian formulation of Kaup-Newell
hierarchy, however here we also give the explicit expressions (2.11b) of the
Hamiltonian functions $H_m$.

Further, similar to Ref. [16], we can get the Zakharov-Shabat equation
$$ V^{(m)}_{t_n}-V^{(n)}_{t_m} +[V^{(m)},V^{(n)}]
=(V^{(m)})'[K_n]-(V^{(n)})'[K_m] +[V^{(m)},V^{(n)}]
=0,$$ which also shows the commutability of the flows of (2.10). 
In addition, we can directly verify  
$$V_{t_m}=[V^{(m)},V],\ m\ge 0,\tag2.12$$
when $u_{t_m}=K_m$, i.e. $U_{t_m}-V_x^{(m)}+[U,V^{(m)}]=0,\ m\ge0$.
In fact, we may deduce that
$V_{t_m}-[V^{(m)},V]$ satisfies the adjoint representation equation
of $\phi _x=U\phi$ and that $(V_{t_m}- [V^{(m)},V])|_{[u]=0}=0$.
Hence (2.12) follows from  the uniqueness property [6] (i.e. if $
V_x=[U,V]$ and $V|_{[u]=0}=0$, then $V$ itself vanishes)
of the adjoint representation equation.
The equality (2.12) shall be used to elucidate the commutability of the
flows generated by nonlinearization in the next section.

\vskip5mm

\newpage 
\centerline{{\bf 3. Binary nonlinearization of Lax pairs}}

\vskip2mm

In this section, we would like to exhibit a binary nonlinearization
procedure, as in Refs. [6,7], for the Lax pairs and adjoint Lax pairs 
of Kaup-Newell hierarchy (2.7) or (2.10). Note that 
$U_{t_m}-V^{(m)}+[U,V^{(m)}]=0$ 
if and only if $(-U^T)_{t_m}-(-(V^{(m)})^T)_x+[-U^T,-(V^{(m)})^T]=0$.
Kaup-Newell systems (2.10) have another kind of Lax pairs called
the adjoint Lax pairs
\vskip2mm
\line {\hbox to 0pt {\hss}
\hss $\displaystyle \left \{\matrix \format \l\\
\psi _{x}=-U^T \psi =-U^T(u,\la )\psi,
\\ \vspace {2mm}
\psi _{t_m}=-(V^{(m)})^T \psi =-(V^{(m)})^T(u,\la )\psi,
\endmatrix  \right. $\hss
\hbox to 0pt {\hss $\displaystyle \matrix (3.1a)\\
\vspace {2mm} (3.1b)\endmatrix $}} \vskip2mm
\noindent where $T$ means the transpose of matrix and $\psi
=(\psi_1,\psi_2)^T$.
The Lax pairs and adjoint Lax pairs are a basic object of binary 
nonlinearization.

\vskip 1mm
\noindent {\bf 3.1. Bargmann symmetry constraint:}

Let us first calculate the variational derivatives of $\la =\la (u)$ 
with respect to
the potentials. Taking Gateaux derivative of
$\phi_x=U(u,\la )\phi$ at $K=(K_1,K_2)^T$ engenders that
$$(\phi'[K])_x=\bigl(\frac {\part U}{\part q}K_1+
\frac {\part U}{\part r}K_2+
\frac {\part U}{\part \la }\la '[K]\bigr)\phi +U\phi '[K].$$
Now by using $\psi_x=-U^T\psi$, we obtain
$$\int _{-\infty}^\infty \psi^T\bigl(\frac {\part U}{\part q}K_1+
\frac {\part U}{\part r}K_2+
\frac {\part U}{\part \la }\la '[K]\bigr)\phi \, dx=0,$$
which leads to $$\la '[K]=\frac1 
{-\int _{-\infty}^\infty <\phi\psi^T,\frac {\part U}{\part \la }>dx}
\int _{-\infty}^\infty\bigl( <\phi\psi^T,\frac {\part U}{\part q
}>K_1+ <\phi\psi^T,\frac {\part U}{\part r}>K_2\bigr)dx.$$
In this way, according to the definition, 
we get the variational derivatives of $\la $ with respect to the
potentials $q,r$:
$$ \frac {\delta \la }{\delta q}=
\frac {<\phi\psi^T,\frac {\part U}{\part q}>}{
-\int _{-\infty}^\infty <\phi\psi^T,\frac {\part U}{\part \la
}>dx},\  \frac {\delta \la }{\delta r}=
\frac {<\phi\psi^T,\frac {\part U}{\part r}>}{
-\int _{-\infty}^\infty <\phi\psi^T,\frac {\part U}{\part \la }>dx}.\tag3.2$$
It follows that
$$ \frac {\delta \la }{\delta u}
=\frac 1 E (\phi_2\psi_1,\la \phi_1\psi_2)^T, \  E=
-\int ^\infty_{-\infty}(\phi _{1}\psi_{1}+r\phi_1\psi_2-\phi_2\psi_1)
\,dx.\tag3.3$$
When the zero boundary conditions:
$\lim_{|x|\to +\infty}\phi=  \lim_{|x|\to +\infty}\psi=0$ are imposed,
we can give out a very useful characterization of
the variational derivatives of $\la $:
$$\Psi \frac {\delta \la }{\delta u}=\la \frac {\delta \la }{\delta u},
\tag3.4$$
where $\Psi$ and $\delta \la /\delta u$ are defined by (2.11a) and (3.3), 
respectively. Introducing $N$ distinct eigenvalues
$\la _1,\la _2,\cdots,\la _N,$ we obtain the following 
spatial and temporal  systems
\vskip3mm
\line {\hbox to 0pt {\hss}
\hss $\displaystyle \left \{\matrix \format \l \\ 
\pmatrix \phi _{1j}\\ \phi _{2j}\endpmatrix 
_x=U(u,\la _j) \pmatrix \phi _{1j}\\ \phi _{2j}\endpmatrix 
,\ j=1,2,\cdots,N,\\ \vspace {3mm}
\pmatrix \psi_{1j}\\ \psi_{2j}\endpmatrix  _x=-U^T(u,\la _j)
\pmatrix \psi_{1j}\\ \psi_{2j}\endpmatrix
,\ j=1,2,\cdots,N; \endmatrix
\right. $\hss \hbox to 0pt {\hss $\displaystyle \matrix (3.5a)\\
\vspace {7.5mm} (3.5b)\endmatrix $}} \vskip2mm
\line {\hbox to 0pt {\hss} \hss $\displaystyle \left \{\matrix \format \l \\ 
\pmatrix \phi _{1j}\\ \phi _{2j}\endpmatrix 
_{t_m}=V^{(m)}(u,\la _j) \pmatrix \phi _{1j}\\ \phi _{2j}\endpmatrix
,\ j=1,2,\cdots,N,\\ \vspace {3mm}
\pmatrix \psi_{1j}\\ \psi_{2j}\endpmatrix  _{t_m}=-(V^{(m)})^T(u,\la _j)
\pmatrix \psi_{1j}\\ \psi_{2j}\endpmatrix 
,\ j=1,2,\cdots,N. \endmatrix \right. $\hss
\hbox to 0pt {\hss $\displaystyle \matrix\ \vspace {-6mm}
\\ \vspace{4mm}(3.6a)\\
\vspace {8.5mm} (3.6b)\\ \ 
\endmatrix $}}
\vskip3mm \noindent
The compatibility condition of (3.5) and (3.6) is still 
the $m$th Kaup-Newell system $u_{t_m}=K_m$.
For (3.5) and (3.6), we may take symmetry constraints 
$$JG_n=J\sum_{j=1}^NE_j \frac {\delta \la _j }{\delta u},\ 
\text {or}\ G_n=\sum_{j=1}^NE_j \frac {\delta \la _j }{\delta u},\ 
E_j= -\int ^\infty_{-\infty}(\phi _{1j}\psi_{1j}+r\phi_{1j}\psi_{2j}-\phi_{2j}
\psi_{1j}) \,dx,\ n\ge0,\tag3.7$$
where $J$ and $G_n,\ n\ge 0,$ are defined by (2.11) and (2.10), respectively.
In this paper, we are interested in the Bargmann constraint (i.e. the first
one in (3.7))
$$JG_0=J\sum_{j=1}^NE_j \frac {\delta \la _j }{\delta u}\ \text{or}\ 
G_0=\sum_{j=1}^NE_j \frac {\delta \la _j }{\delta u}.\tag3.8$$
This kind of constraints  requires the $G$-vector fields to be
a potential linear function not including any potential differential.
The constraint (3.8) yields that $$\pmatrix c_0\\b_0\endpmatrix =\bigl (
<P_2,Q_1>, <AP_1,Q_2>\bigr)^T,$$
which gives rise to  the following binary nonlinear constraint between the
potential and the eigenfunctions and adjoint eigenfunctions
$$q= <AP_1,Q_2>,\  r= <P_2,Q_1>.\tag3.9$$
Here for economic writing, we have used the notation: 
$P_i=(\phi_{i1},\cdots,\phi_{iN})^T,
\ Q_i=(\psi_{i1},\cdots,\psi_{iN})^T,$ $i=1,2$, 
$A=\text{diag}(\la _1,\la _2,\cdots, \la _N)$,
$<y,z>=\sum
_{j=1}^Ny_jz_j$ for two $N$-dimensional vectors 
$y=(y_1,\cdots,y_N)^T,\,z=(z_1,\cdots,z_N)^
T\in R^N$.

We point out that the above explanatory deduction requires 
$E_j\ne 0,\ 1\le j\le N$. But in fact, what we need below 
is just the binary nonlinear
constraint (3.9). It itself doesn't touch the condition: 
$E_j\ne 0,\ 1\le j\le N$.

\vskip 1mm
\noindent {\bf 3.2. Nonlinearized Lax pairs and adjoint Lax pairs:}

In what follows,
 we will denote by $\widetilde  P$ the expression of $P(u)$ under the
constraint (3.9). 
Having substituted (3.9) into the Lax pairs and the adjoint Lax pairs: 
(3.5) and (3.6), we acquire the nonlinearized Lax pairs and adjoint Lax pairs
\vskip3mm
\line {\hbox to 0pt {\hss}
\hss $\displaystyle \left \{\matrix \format \l \\ 
\pmatrix \phi _{1j}\\ \phi _{2j}\endpmatrix 
_x= \pmatrix \la _j&\widetilde q\\ \vspace{1mm}
\la _j\widetilde r&-\la _j\endpmatrix 
\pmatrix \phi _{1j}\\ \phi _{2j}\endpmatrix 
,\ j=1,2,\cdots,N,\\ \vspace {3mm}
\pmatrix \psi_{1j}\\ \psi_{2j}\endpmatrix 
_x=\pmatrix -\la _j &-\la _j\widetilde r \\ \vspace{1mm}
-\widetilde q&\la _j\endpmatrix 
\pmatrix \psi_{1j}\\ \psi_{2j}\endpmatrix  ,\ j=1,2,\cdots,N;
\endmatrix \right. $\hss
\hbox to 0pt {\hss $\displaystyle \matrix (3.10a)\\
\vspace {8.5mm} (3.10b)\endmatrix $}}
\vskip2mm
\line {\hbox to 0pt {\hss}
\hss $\displaystyle \left \{\matrix \format \l \\ 
\pmatrix \phi _{1j}\\ \phi _{2j}\endpmatrix 
_{t_m}=\pmatrix \sum_{i=0}^m \widetilde a_i \la _j^{m+1-i}&
\sum_{i=0}^m \widetilde b_i \la _j^{m-i}\\ \vspace{1mm}
\sum_{i=0}^m \widetilde c_i \la _j^{m+1-i}&
-\sum_{i=0}^m \widetilde a_i \la _j^{m+1-i}\endpmatrix
\pmatrix \phi _{1j}\\ \phi _{2j}\endpmatrix
,\ j=1,2,\cdots,N,\\ \vspace {3mm}
\pmatrix \psi_{1j}\\ \psi_{2j}\endpmatrix 
_{t_m} =\pmatrix -\sum_{i=0}^m \widetilde a_i \la _j^{m+1-i}&
-\sum_{i=0}^m \widetilde c_i \la _j^{m+1-i}\\ \vspace{1mm}
-\sum_{i=0}^m \widetilde b_i \la _j^{m-i}&
\sum_{i=0}^m \widetilde a_i \la _j^{m+1-i}\endpmatrix
\pmatrix \phi _{1j}\\ \phi _{2j}\endpmatrix
,\ j=1,2,\cdots,N. \endmatrix \right.\quad $\hss
\hbox to 0pt {\hss $\displaystyle \matrix \ \vspace{-3mm}\\ (3.11a)\\
\vspace {9.5mm} (3.11b)\\ \ 
\endmatrix $}}
\vskip3mm
\noindent We easily see that the spatial part of the nonlinearized 
Lax pairs and adjoint Lax pairs, namely, the system (3.10)
is a {\it finite dimensional system}  with regard to $x$, but
the temporal  parts of the nonlinearized Lax pairs and adjoint Lax pairs, 
namely,
the systems (3.11) for $n\ge 0$ are all systems of {\it evolution equations}
with regard to $t_m,x$. 
Obviously the system (3.10)  reads 
$$\left \{ \aligned &  P_{1x}=AP_1+<AP_1,Q_2>P_2,\\
&P_{2x}=<P_2,Q_1>AP_1-AP_2,\\
&Q_{1x}=-AQ_1-<P_2,Q_1>AQ_2,\\
&Q_{2x}=-<AP_1,Q_2>Q_1+AQ_2,\endaligned \right.$$
and it may be represented as the following Hamiltonian form
$$P_{ix}=\frac {\part H}{\part Q_{i}},\ Q_{ix}=-\frac {\part
H}{\part P_i},\ i=1,2,\tag3.12a$$
where the Hamiltonian function  
$$ H=<AP_1,Q_1>-<AP_2,Q_2>+
<P_2,Q_1> <AP _1,Q_2> .\tag3.12b$$
In the following, we shall prove that the 
system (3.10) is an integrable  finite dimensional 
Hamiltonian system in the Liouville sense [17] and that under
the control of the system (3.10), the systems (3.11) for $m\ge 0$ are all
transformed into integrable finite dimensional Hamiltonian systems 
in the Liouville sense [17].

\vskip 1mm
\noindent {\bf 3.3 An involutive and independent system:}

To show the Liouville 
integrability of the nonlinearized systems 
(3.10) and (3.11), we need an involutive and independent system
to generate integrals of motion for them. We will use the nonlinearized 
Lax pairs and adjoint Lax pairs in the
case of zero 
boundary condition on the eigenfunctions and adjoint eigenfunctions
in order to derive such an involutive and 
independent system. Under the zero 
boundary condition,
the characterization (3.4) asserts that 
$$\pmatrix \widetilde  c_m \\ \vspace {2mm}
\widetilde b _m \endpmatrix =\widetilde  \Psi ^{m}
\pmatrix \widetilde  c_0 \\ \vspace {2mm}
\widetilde  b _0 \endpmatrix  =\pmatrix <A^mP_2,Q_1>  \\
\vspace {2mm} <A^{m+1}P_1,Q_2>\endpmatrix,\ m\ge 0,\tag3.13a$$
and further from (2.3), we obtain
$$ \widetilde  a_0=1,\  \widetilde  a_m
=\part ^{-1}(\widetilde  q \,\widetilde  c_m-\widetilde
r\,\widetilde  b_m)=\frac12 \bigl(<A^mP_1,Q_1>-<A^mP_2,Q_2>\bigr)
,\ m\ge 1.\tag3.13b$$
It is easy to find
 that $\widetilde  V_x=[\widetilde  U,\widetilde V]$ still 
holds.
Thus an obvious equality $(\widetilde  V^2)_x=[\widetilde U,\widetilde  V^2]$
leads to  $$F_x=(\frac12 \text{tr}\widetilde  V^2)_x=\frac d {dx}(
\widetilde  a^2+\la ^{-1} \widetilde  b \widetilde  c)=0.$$
This allows us to conclude that $F$ is a generating function of
integrals of motion for (3.12). Due to that $F=\sum_{n\ge 0}F_n\la ^{-n}$,
we obtain the following expressions of integrals of motion 
$$F_n=\sum_{i=0}^n ( \widetilde  a_i\widetilde  a_{n-i}+\sum _{i=0}^{n-1} 
\widetilde  c_{i} \widetilde  b_{n-1-i}) .$$
Substitution (3.13) into the above equality gives us the explicit 
formulas of $F_n$: $$\align F_0=&1,\ F_1=
<AP_1,Q_1>-<AP_2,Q_2>+ <P_2,Q_1>
<AP _1,Q_2>=H, \tag3.14a\\
F_n=& \sum_{i=1}^{n-1}\frac14 \big(
<A^{i}P _1,Q_1>-<A^iP_2,Q_2>\bigr)\bigl(
<A^{n-i}P _1,Q_1>-<A^{n-i}P_2,Q_2>\bigr)
\\ \vspace {1mm}&+ \sum _{i=0}^{n-1}<A^iP_2,Q_1><A^{n-i}P_1,Q_2>
+<A^nP_1,Q_1>-<A^nP_2,Q_2>
,\ n\ge2.\qquad\tag3.14b\endalign$$
Note that we still have a similar equality $\widetilde V_{t_m}=[\widetilde 
V^{(m)}, \widetilde V],\,m\ge 0,$ to (2.12). With the same argument,
we may find that $F=\frac12 \text{tr}\widetilde  V^2$ is also a generating
function of integrals of motion for (3.11). Moreover,
a direct calculation can give rise to 
$$Q_{it_m}=-\frac {\part ( F_{m+1})}{\part P_i},\  
P_{it_m}=\frac {\part ( F_{m+1})}{\part Q_i},\  i=1,2. \tag3.15 $$
Actually, we have  $$\bigl ( \frac {\part F}{\part P_i},
\frac {\part F}{\part Q_i}\bigr)^T =\Bigl (
\text{tr} \big (\widetilde  V\frac \part {\part P_i}\widetilde  V\bigr),
\text{tr} \big (\widetilde  V\frac \part
{\part Q_i}\widetilde  V\bigr)\Bigr)^T,\ i=1,2.\tag3.16$$
Further it is carried out that $$\align &
 \text{tr} \big (\widetilde  V\frac \part {\part P_1}\widetilde  V\bigr)=
\text{tr}\sum_{i=0}^\infty  \pmatrix \widetilde  a_i &\widetilde  b_{i-1}\\
\widetilde  c_i&-\widetilde  a_i\endpmatrix\la ^{-i}
\sum_{j=0}^\infty\frac \part{\part P_1} 
\pmatrix \widetilde  a_j &\widetilde  b_{j-1}\\
\widetilde  c_j&-\widetilde  a_j\endpmatrix\la ^{-j}
\\&=\text{tr}\sum _{\Sb i\ge 0\\ \vspace {1mm}j\ge1\endSb}
\pmatrix \widetilde  a_i &\widetilde  b_{i-1}\\
\widetilde  c_i&-\widetilde  a_i\endpmatrix
\pmatrix \frac12A^{j}Q_1&A^{j}Q_2\\0&-\frac12A^{j}Q_1\endpmatrix
\la ^{-(i+j)}\\&= \sum _{\Sb i\ge 0\\ \vspace {1mm}j\ge1\endSb}
\bigl ( \widetilde  a_i A^{j}Q_1+\widetilde  c_i 
A^{j}Q_2\bigr) \la ^{-(i+j)}.\tag3.17\endalign $$
Similarly we can give
$$ \text{tr} \big
(\widetilde  V\frac \part {\part P_2}\widetilde  V\bigr)=
\bigl(\sum _{\Sb i\ge 1\\ \vspace {1mm}j\ge0\endSb}
\widetilde  b_{i-1} A^{j}Q_1
-\sum _{\Sb i\ge 0\\ \vspace {1mm}j\ge1\endSb}
\widetilde  a_i A^{j}Q_2\bigr) \la ^{-(i+j)}.\tag3.18$$
It follows from (3.16), (3.17) and (3.18) that 
$$\bigl( \dfrac {\part F_{m+1}}{\part P_1},
\dfrac {\part F_{m+1}}{\part P_2}\bigr)^T =-(\widetilde 
V^{(m)})^T\pmatrix Q_1\\ Q_2\endpmatrix,\ m\ge0.$$
A complete similar argument can lead to
$$\bigl( \dfrac {\part F_{m+1}}{\part Q_1},
\dfrac {\part F_{m+1}}{\part Q_2}\bigr)^T =\widetilde 
V^{(m)}\pmatrix P_1\\ P_2\endpmatrix,\ m\ge0.$$
The above two equalities enables us to write all the flows of (3.11)
in the Hamiltonian form (3.15).

Now we can easily show the involution 
$F_n,\ n\ge 1$.
It is known that for the standard symplectic structure on $R^{4N}$
$$\omega ^2=dQ_1\wedge dP_1+dQ_2\wedge dP_2,\tag3.19$$
the corresponding Poisson bracket is defined as follows
$$\{f,g\}=\omega ^2 (Idf,Idg),\tag3.20$$
where $Idh$ expresses a Hamiltonian vector field of a smooth function 
$h$ on $R^{4N}$, defined by $Idh\,\text{\_\kern-.1em l}\,\omega^2=-dh$,
with $\text{\_\kern-.1em l}$ being the left interior product.
Because $F$ is also a generating function of integrals of
motion for (3.11), it is found that 
$$\{F_{n+1},F_{m+1}\}=\frac \part {\part t_m}F_{n+1}=0,\ m,n\ge 0,\tag3.21$$
by using the Hamiltonian structure (3.15) of (3.11).
This implies that $\{F_n\}_{n=0}^\infty$ constitutes an involutive system
with regard to (3.20). It may also directly be shown that 
 $\{F_m,F_n\}=0,\ m,n\ge1$.

\proclaim{Theorem 3.1}{\bf Theorem 3.1:} Let $\bar F_k,\ 1\le k\le N,$ be 
defined by 
$$ \bar F_k=\phi_{1k}\psi_{1k}+\phi_{2k}\psi_{2k},\ 1\le k\le N. $$
Then $\bar F_{k},\ 1\le k\le N, \ F_n,\ n\ge 1$,
constitute an involutive system, of which  
$\bar F_{k}, \ F_n,\ 1\le k, n\le N,$
are functionally independent on some region of $R  ^{4N}$.
\endproclaim
\demo {Proof} {\bf Proof:}
Noting that we already have $\{ F_m,F_n\}=0,\  m, n\ge 1,$
it suffices for the involution of $\bar F_k,\ 1\le k\le N,\ F_n,\ n\ge 1,$
that we prove
 $\{\bar F_k,F_n\}=0,\ 1\le k\le N,\ n\ge 1.$
This needs just a simple calculation due to the particular form of
$\bar F_k,\ 1\le k\le N$.

In what follows, we show that there exists some region of $R  ^{4N}$ over 
which the functions $\bar F_k,\ F_n,\ 1\le k,n\le N,$
can become an  independent system. 
Suppose that this result is not true. Namely,
there doesn't exist any region of $R ^{4N}$ over which 
the functions $\bar F _k,\  F_n,\ 1\le k,n\le N,$ could
be independent. Hence there exist $2N$ constants $\alpha _k$,
$ \beta  _n,\ 1\le k,n\le N,$ satisfying 
$$ \sum_{k=1}^N\alpha _k^2 +\sum_{n=1}^{N}\beta ^2 _n\ne 0,$$ 
so that we have for all points $(P_1^T,P_2^T,Q_1^T,Q_2^T)$ in $ R ^{4N}$
$$ \sum_{k=1} ^N\alpha _k\left ( \left (\frac {\part \bar F _k}{ \part P_1}
\right)^T,
\left (\frac {\part \bar F _k}{ \part P_2}\right)^T
\right) +\sum_{n=1}^{N}\beta  _n\left (\left (\frac {\part  F _n}
{ \part P_1}\right)^T, 
\left (\frac {\part  F _n}{ \part P_2}\right)^T
\right)=0.\tag3.22$$ 
From the above equality, we would like to derive that all constants
$\alpha _k$,
$\beta  _n,\ 1\le k,n\le N,$ must be zero. 
 We will utilize the following equalities:
$$ \aligned & \left.\frac {\part \bar F _k}{ \part P_i}\right|_{P_1=P_2=0}
={(\undersetbrace k-1 \to {0,\cdots,0},\psi _{ik},\undersetbrace N-k \to{
0,\cdots,0})}{}^T,\ 
i=1,2,\ 1\le k\le N,\\
&\left.\frac {\part F_1}{\part P_1}\right|_{P_1=P_2=0}=AQ_1,\ 
\left.\frac {\part F_1}{\part P_2}\right|_{P_1=P_2=0}=-AQ_2, \\
& \left.\frac {\part  F _n}{ \part P_1}\right|_{P_1=P_2=0}
= A^nQ_1,\ 
\left. \frac {\part  F _n}{ \part P_2}\right|_{P_1=P_2=0}
=-A^nQ_2,\ n\ge 2
.\endaligned
$$
The equality (3.22) upon choosing $P_1=P_2=0$
 engenders
$$ \alpha _k=\sum_{n=1}^N\beta _n\la _k^n=
-\sum_{n=1}^N\beta _n\la _k^n,\ 1\le k\le N.$$
Therefore we have $\sum_{n=1}^N\beta _n\la _k^n=0,\ 1\le k\le N.$
Let $f(\la )=\sum_{n=1}^{N}\beta  _n\la  ^{n}$, which is a polynomial 
of degree $N$. Since it possesses $N$ distinct
roots $\la _1,\la _2,\cdots,\la _N$, we have $f(\la )=0$. This means that
$\beta _n=0,\ 1\le n\le N,$ and further
we obtain $\alpha _k=0,\ 1\le k\le N$. In this way,
 we arrive at all zero constants,
which contradicts to our assumption.  
Hence there must exists at least a region $\Omega \subseteq R ^{4N}$ 
over which the
$2N$ 1-forms $d\bar F_k,\,   dF_{n},\  1\le k,n\le N,$ are  linearly 
independent. The proof is finished.
\enddemo

The theorem provides us with an involutive and independent system of 
polynomials, which is completely new as far as we know. 

\vskip 1mm
\noindent {\bf 3.4 Liouville integrability of the nonlinearized Lax pairs
and adjoint Lax pairs:}

We now proceed to show the Liouville integrability of the nonlinearized 
Lax pairs and adjoint Lax pairs, i.e. (3.10) and (3.11).
We point out that the above manipulation with the explicit expression (3.15)
 of (3.11)
requires the zero boundary conditions:
$\lim_{|x|\to +\infty}Q_i= 
\lim_{|x|\to +\infty}P_i=0,$ $i=1,2$. Now let us deal with a
general case, i.e. the case where 
the zero boundary conditions are not imposed.
In this case, it follows from 
$\phi_x=U(u,\la )\phi$,
$\psi_x=-U(u,\la )^T\psi$ that
$$\Psi \pmatrix \phi _2 \psi_1\\ \la \phi _1\psi_2\endpmatrix=\la 
\pmatrix \phi _2 \psi_1\\ \la \phi _1\psi_2\endpmatrix
+I\pmatrix r \\ q \endpmatrix,
\tag3.23$$
where $I$ is an integral of motion of 
$\phi_x=U(u,\la )\phi$,
$\psi_x=-U(u,\la )^T\psi$.
By applying (3.23) $n$ times, we easily work out that
$$\pmatrix \widetilde c_{m}\\ \vspace{1mm}
\widetilde b_{m}\endpmatrix =\widetilde
\Psi^m \pmatrix \widetilde c_0 \\ \vspace{1mm}\widetilde b_0 \endpmatrix 
=\sum_{i=0}^mI_i
 \pmatrix <A^{m-i}P _2,Q _1>\\ \vspace{1mm}
<A^{m+1-i}P _1,Q _2>\endpmatrix ,\
 m\ge 0, \tag3.24$$
where $I_0=1$ and $I_{i},\, 1\le i\le m,$ are all integrals of motion of
(3.10).
Further we can compute that
$$\align &\widetilde a_{m}=\part ^{-1}(\widetilde q 
\widetilde c_{m}-\widetilde r \widetilde b_{m})\\
=&\frac12 \sum_{i=0}^{m-1}I_i(<A^{m-i}P_1,Q_1>
-<A^{m-i}P_2,Q_2>)+T_{m},\
m\ge1,\quad\tag3.25\endalign$$
where $T_{m}$ is an integral of motion of (3.10), too.
The latter two equalities in (2.3) lead to that $T_{m}=-I_m,\ m\ge1$. 
Therefore by using (2.5), we obtain 
$$\align
&\sum_{i=0}^{m-1}I_i(<A^{m-i}P_1,Q_1>-<A^{m-i}P_2,Q_2>)-2I_m\\
=&-\sum_{i=1}^{m-1}\frac14
\Bigl[\sum_{k=0}^{i-1}I_k(<A^{i-k}P_1,Q_1>-
<A^{i-k}P_2,Q_2>)-2I_i\Bigr]\times\\&
\Bigl[\sum_{l=0}^{m-i-1}I_l(<A^{m-i-l}P_1,Q_1>-
<A^{m-i-l}P_2,Q_2>)-2I_{m-i}\Bigr]\\&
-\sum_{i=1}^{m-1}
\sum_{k=0}^{i}I_k<A^{i+1-k}P_1,Q_2>\sum_{l=0}^{m-1-i}I_l
<A^{m-1-i-l}P_2,Q_1>,\ m\ge2.\tag3.26\endalign$$
By interchanging the summing in the above equality and noting the
expressions of $F_n'$s,
a direct but quite lengthy calculation may give rise to
$$I_m=\frac 12 \sum \Sb k+l\le m-1\\ \vspace {1mm}k,l\ge0\endSb
I_kI_lF_{m-(k+l)}-\frac12 \sum _{i=1}^{m-1}I_iI_{m-i},\
m\ge2.\tag3.27$$
This equality and $I_1=\frac12 F_1$, determined
by $\widetilde a_1=-\frac12
\widetilde q\widetilde r$, readily yield the following expressions of $I_m'$s:
$$I_m=\sum _{n=1}^md_n\sum\Sb i_1+\cdots+i_n=m\\i_1,\cdots,i_n\ge
1\endSb 
F_{i_1}\cdots
F_{i_n},\ m\ge1.\tag3.28$$
Here the constants $d_n$ are defined recursively by 
$$d_1=\frac 12,\ d_2=\frac38,\ d_n=d_{n-1}+\frac12
\sum_{s=1}^{n-2}d_sd_{n-s-1}-\frac12 \sum_{s=1}^{n-1}d_sd_{n-s},\
n\ge3.\tag3.29$$
At this point, we can give the expressions of $\widetilde a_m,\,\widetilde
b_m,\,\widetilde c_m$ in terms of $F_n$ and further a direct
computation may show that
the temporal  parts (3.11) of the nonlinearized Lax pairs and
adjoint Lax pairs are represented as the following Hamiltonian
systems
$$ Q _{it_m}=-\frac {\part H_{m}}{\part P _i},\ 
P _{it_m}=\frac {\part H_{m}}{\part Q _i},
\ i=1,2\tag3.30a$$
with the Hamiltonian functions
$$H_m= 
\sum _{n=0}^m \frac {d_n}{n+1}
\sum\Sb i_1+\cdots+i_{n+1}=m+1\\ i_1,\cdots,i_{n+1}\ge1\endSb
F_{i_1}
\cdots F_{i_{n+1}}\ (d_0=1),\ 
m\ge 0.\tag3.30b$$
The Hamiltonian phase flows $g^{t_m}_{H_m}$ of the systems (3.30)
commute with each other due to the commutability of $F_n$. 

Because  $H,\ 
H_m,\ m\ge 1,$  are all functions of $F_n,\ n\ge 1,$ 
we can now conclude the Liouville integrability on the nonlinearized 
Lax pairs and adjoint Lax pairs.

\proclaim{Theorem 3.2}{\bf Theorem 3.2:}
The spatial part (3.10) of  the nonlinearized Lax pairs and 
adjoint Lax pairs and 
the temporal  parts (3.11) of the nonlinearized Lax pairs and adjoint 
Lax pairs 
under the control of the spatial part are all finite dimensional
Liouville integrable Hamiltonian systems with the corresponding
 Hamiltonian 
functions $H$ and $H_m$ defined by (3.13b) and (3.30b).
Moreover they all possess involutive and independent integrals of  motion:
$\bar F_k,\ F_n,\ 1\le k,n\le N$.
\endproclaim

\vskip 1mm
\noindent {\bf 3.5. Involutive solutions to Kaup-Newell hierarchy:}

The above manipulation also allows us to establish 
the following involutive solutions of
the $m$th Kaup-Newell system 
$u_{t_m}=K_m$:
\vskip3mm
\line {\hbox to 0pt {\hss}
\hss $\displaystyle \left \{\matrix 
q(x,t_m)=<A g^x_Hg^{t_m}_{H_{m}}P_1(0,0),
g^x_Hg^{t_m}_{H_{m}}Q_2(0,0)>\\
\vspace {2mm}
r(x,t_m)=
<g^x_Hg^{t_m}_{H_{m}}P_1(0,0),
g^x_Hg^{t_m}_{H_{m}}Q_2(0,0)>\ \ 
\endmatrix 
\right. $\hss
\hbox to 0pt {\hss $\displaystyle \matrix  
(3.31a)\\
\vspace {3mm}(3.31b)\endmatrix $}}
\vskip3mm
\noindent where $g^x_H,\,g^{t_m}_{H_{m}}$ are the Hamiltonian phase
flows [17] associated with the Hamiltonian systems (3.12) and
(3.30), respectively, but
$P_i(0,0),\,Q_i(0,0),\, i=1,2$,
may be arbitrary.
This provides us with a way to present solutions
to Kaup-Newell systems by solving two separate finite dimensional
integrable systems with the variables $x$ and $t_m$. 
In fact, 
in view of the Liouville integrability of the nonlinearized Lax pairs
and adjoint Lax pairs,
this kind of involutive solutions to integrable
systems not only gives rise to the interrelation between $1+1$
dimensional integrable systems and finite dimensional integrable
systems, but also exhibits the characteristic of integrability by quadratures 
for integrable systems in $1+1$ dimensions.

\vskip5mm

\centerline{{\bf 4. Conclusions and remarks}}

\vskip2mm

We have introduced a different spectral problem from the original one
for Kaup-Newell systems and discussed the corresponding
Bargmann symmetry constraint. 
We have also 
exhibited a new explicit involutive system 
$ \{ \bar F_k,\,1\le k\le N,\,
F_n,\,n\ge1\}$
on the symplectic manifold $(R ^{4N},\omega ^2)$, of which 
 $ \{ \bar F_k,\,
F_n,\,1\le k,n\le N\}$ constitute an independent system.
The Lax pairs and adjoint Lax pairs are nonlinearized into a hierarchy of 
finite dimensional Liouville integrable Hamiltonian systems and
 Bargmann symmetry constraint leads to 
 a sort of  involutive
 solutions to Kaup-Newell systems, which may be presented by solving
two commutative finite dimensional Hamiltonian systems, i.e.
 the spatial part and 
the temporal part of
nonlinearized Lax pairs and adjoint Lax pairs.

We remark that 
under the reduction $Q_1=-P_2, \,Q_2=P_1$, the involutive system
$\{F_n\}_{n=1}^\infty$ defined
by (3.14) are reduced to 
$$\align F_1=&
-2<AP_1,P_2>-<AP_1,P_1><P_2,P_2>,\\
F_n=&\sum_{i=0}^n\Bigl(<A^iP_1,P_2><A^{n-i}P_1,P_2>
-<A^iP_1,P_1><A^{n-i}P_2,P_2>\Bigr)\\
&-2<A^nP_1,P_2>-<P_1,P_2><A^nP_1,P_2>+
<P_1,P_1><A^nP_2,P_2>,\ n\ge2.\endalign$$
This involutive system 
corresponds to mono-nonlinearization for the
Kaup-Newell systems (2.10) (see Ref. [18]).
It involves $2N$ dependent variables but not $4N$ dependent variables.

It should be noted that  binary nonlinearization
procedure is theoretically very systematic (refer to [19] 
for more information). 
It paves a way to show Liouville integrability for both the spatial part
(sometimes called constrained flows or 
restricted flows) and the temporal parts 
of nonlinearized Lax pairs and adjoint Lax pairs, which is short of discussion
in the works of restricted flows (see for instance [8]). We only need 
the distinction of $\la _1,\la _2,\cdots,\la _N$ for proving
Liouville integrability but may have the condition
$\prod_{i=1}^N\la _i=0$, which is not allowed in the case of restricted flows. 
The resulting binary nonlinear constraints involve two classes of
eigenfunctions and thus might greatly extend explicit finite
dimensional integrable Hamiltonian systems in the Liouville sense. 

\vskip 0.2cm
\noindent {\bf 
Acknowledgment:} 
The authors would like to thank
the National Natural Science Foundation of China (Grant 19301015)
and the Shanghai Science and Technology Commission
of China for financial support. One of the authors (W. X. Ma) 
is also very grateful to 
the Alexander von Humboldt Foundation for a research fellow
and International Mathematical Union
for supporting him to
communicate on some results of the paper at ICM 1994, Z\"urich.

\vskip 0.2cm

\centerline{References}
\vskip 0.2cm

\item {[1]} Ablowitz M. J., Ramani A. and Segur H., {\it Lett. Nuovo
Cim.}, {\bf 23} (1978) 333; {\it J. Math. Phys.} {\bf 21} (1980) 715, 1006.
\item {[2]} Airault H., McKean H. P. and Moser J., {\it Commun. Pure
Appl. Math.}, {\bf 30} (1977) 95; Case K. M. 
{\it Proc. Nat'l. Acad. Sci.},
{\bf 75} (1978) 3562;  
{\it Proc. Nat'l. Acad. Sci.}, {\bf 76} (1979) 1.
\item {[3]}
Cao C. W., {\it Chin. Quart. J. Math.}, {\bf 3} (1988) 90; 
{\it Sci. China A}, {\bf 33} (1990)  528.
\item {[4]} Cao C. W. and Geng X. G.,
 {\it J. Phys. A: Math. Gen.},
{\bf 23}  (1990)  4117;  {\it J. Math. Phys.}, {\bf 32}  (1991) 2323.
\item {[5]}
Zeng Y. B. and Li Y. S., {\it J. Math. Phys.}, {\bf 30} (1989) 1679;
Zeng Y. B.,  {\it Phys. Lett. A}, {\bf 160} (1991) 541.
\item {[6]} Ma W. X. and Strampp W.,
{\it Phys. Lett. A}, {\bf 185} (1994) 277. 
\item {[7]} Ma W. X., 
{\it J. Phys. Soc. Jpn.}, {\bf 64} (1995) 1085; {\it Physica A},
{\bf 219} (1995) 467; Binary
nonlinearization for the Dirac systems, to appear in {\it
Chinese Annals of Math. B}
(solv-int/9512002).
\item {[8]}
Antonowicz M. and Wojciechowski S.,  
{\it Phys. Lett. A}, {\bf 147} (1990) 455;
 {\it J. Phys A: Math. Gen.}, {\bf 24} (1991) 5043;
 {\it J. Math. Phys.}, {\bf 33} (1992) 2115;
Ragnisco O. and Wojciechowski S.,  
{\it Inverse Problems}, {\bf 8} (1992) 245; 
Blaszak M.,
{\it Phys. Lett. A}, {\bf 174} (1993) 85; Tondo G., 
{\it J. Phys. A: Math. Gen.},  {\bf 28} (1995) 5097.
\item {[9]}
Kaup D. J. and Newell A. C.,
{\it J. Math. Phys.}, {\bf 19} (1978) 798.
\item {[10]}
Tu G. Z.,  {\it  J. Phys. A: Math. Gen.},
{\bf 22} (1989) 2375; Ma W. X., {\it J. Phys. A: 
Math. Gen.}, {\bf 26} (1993) 2573.
\item {[11]}
Fordy A. P. and Gibbons J.,  {\it J. Math. Phys.}, {\bf 21} (1980) 2508;
 ibid, {\bf 22} (1981) 1170.
\item {[12]}
Fuchssteiner B.,  {\it Nonlinear Anal. TMA}, {\bf 3} (1979)
849;  {\it Prog. Theor. Phys.}, {\bf 65} (1981) 861.
\item {[13]}
Tu G. Z.,  {\it Scientia Sinica A}, {\bf 24} (1986) 138;
{\it J. Math. Phys.}, {\bf 30} (1989) 330.
\item {[14]} Tu G. Z. and Meng D. Z.,  {\it Nonlinear Evolutions}, ed.
Leon J. J. P. (World Scientific, Singapore) 1988 p425; 
 {\it Acta Math. Appl. Sinica}, {\bf 5} (1989) 89.
\item {[15]} Magri F., {\it Nonlinear Evolution Equations and
Dynamical Systems}, eds. Boiti M., Pempinelli F. and Soliani G .
(Springer-Verlag, Berlin) 1980 p233
\item {[16]}
Ma W. X., {\it J. Math. Phys.}, {\bf 33} (1992) 2464;  {\it 
J. Phys. A: Math. Gen.}, {\bf 25} (1992) 5329.
\item {[17]} Arnold V. I., {\it Mathematical Methods of Classical
Mechanics} (Springer-Verlag, Berlin) 1978; 
Abraham R.   and  Marsden J.,  {\it Foundations of Mechanics}, 2nd edition
(Addison-Wesley, Massachusetts, Reading) 1978.
\item {[18]} Liu C. P.,  {\it Appl. Math.--J. Chin. Universities}
 Ser. A, {\bf
8} (1993) 157.
\item {[19]} Ma W. X. and Fuchssteiner B., 
Binary nonlinearization of Lax pairs, to appear in 
{\it Proceedings of International Conference of Nonlinear Physics, Gallipoli,
Italy}
(World Scientific, Singapore).
\bye